\documentclass[letter]{aa}
\usepackage[varg]{txfonts}

\usepackage{natbib}
\usepackage{hyperref}
\usepackage{amsmath}
\hypersetup{colorlinks=true,citecolor=blue}
\usepackage{graphicx}
\usepackage{url}
\usepackage{amsmath}
\usepackage{amssymb} 
\usepackage{amsfonts}
\usepackage{aas_macros}

\title{PRISM: Recovery of the primordial spectrum from Planck data}
\author{ F. Lanusse \inst{1} \thanks{francois.lanusse@cea.fr} \and P. Paykari \inst{1} \thanks{paniez.paykari@cea.fr} \and  J.-L. Starck \inst{1} \and  F. Sureau \inst{1} \and  J. Bobin \inst{1} \and A. Rassat \inst{2}}
\institute{$^1$ Laboratoire AIM, UMR CEA-CNRS-Paris 7, Irfu, SAp, CEA Saclay, F-91191 Gif sur Yvette cedex, France.\\$^2$ Laboratoire d'Astrophysique, Ecole Polytechnique F\'ed\'erale de Lausanne (EPFL), Observatoire de Sauverny, CH-1290, Versoix, Switzerland.}

\begin{document}
\label{firstpage}
\date{\today}

\abstract
{}
{
The primordial power spectrum describes the initial perturbations that seeded the large-scale structure we observe today. It provides an indirect probe of inflation or other structure-formation mechanisms. In this letter, we recover the primordial power spectrum from the Planck PR1 dataset, using our recently published algorithm \textbf{PRISM}.
}
{
PRISM is a sparsity-based inversion method, that aims at recovering features in the primordial power spectrum from the empirical power spectrum of the cosmic microwave background (CMB). This ill-posed inverse problem is regularised using a sparsity prior on features in the primordial power spectrum in a wavelet dictionary. Although this non-parametric method does not assume a strong prior on the shape of the primordial power spectrum, it is able to recover both its general shape and localised features. As a results, this approach presents a reliable way of detecting deviations from the currently favoured scale-invariant spectrum.
}
{
We applied PRISM to 100 simulated Planck data to investigate its performance on Planck-like data. We then applied PRISM to the Planck PR1 power spectrum to recover the primordial power spectrum. We also tested the algorithm's ability to recover a small localised feature at $k \sim 0.125$ Mpc$^{-1}$, which caused a large dip at $\ell \sim 1800$ in the angular power spectrum.
}
{
We find no significant departures from the fiducial Planck PR1 near scale-invariant primordial power spectrum with $A_s=2.215\times10^{-9}$ and $n_s = 0.9624$.
}
\keywords{Cosmology : Primordial Power Spectrum, Methods : Data Analysis, Methods : Statistical}

\maketitle

\section{Introduction}
\label{sec:intro}

The primordial power spectrum describes the initial curvature perturbations that over time evolved to form the large-scale structure we observe today. Because the physics of the early Universe are encoded in the primordial power spectrum, it represents an invaluable probe of primordial cosmology, and measuring it is a crucial research area in modern cosmology. The currently favoured model describing the physics of the early Universe, inflation \citep{guth,Linde-inflation1982}, produces initial perturbations from quantum fluctuations during an epoch of accelerated exponential expansion. This inflation process produces a power spectrum of specific shape and can leave characteristic features. For the simplest inflation models, the power spectrum, generated by almost purely adiabatic perturbations, which is predicted to be nearly scale invariant. Hence, it is often expressed in terms of an amplitude $A_s$ and a spectral index $n_s$ with an optional `running' $\alpha_s$,
\begin{equation}
P(k)=A_s \left(\frac{k}{k_p}\right)^{n_s-1+\frac{1}{2}\alpha_s\ln \left({k}/{k_p}\right)}\;,
\label{eq:pkdef2}
\end{equation}
where $k_p$ is a pivot scale. We consider here only the first-order expansion of the spectral index, although higher orders can be considered \cite[e.g.,][]{Rassat:2010}. Exact scale invariance, known as the Harrison-Zeldovich (HZ) model, which sets $n_s=1$ (and $\alpha_s=0$) \citep{Harrison1970,Zeldovich1972}, has been ruled out by different datasets. Instead, the near scale-invariant spectrum with $n_s<1$ fits the current observations very well \citep[e.g.,][]{PlanckCP}. There exist more complex models that generate deviations from scale invariance. For a short review of inflation we refer to \citet{PlanckPk}.

The recent Planck mission cosmic microwave background (CMB) temperature anisotropy data, combined with the WMAP large-scale polarisation, constrain the spectral index to $n_{s}=0.9603 \pm0.0073$ \citep{PlanckCP}, ruling out exact scale invariance at over $5 \sigma$. Planck also failed to find a statistically significant running of the scalar spectral index, obtaining $\alpha_s =-0.0134\pm0.0090$. On the other hand, high-resolution CMB experiments, such as the South Pole Telescope (SPT)\footnote{\url{http://pole.uchicago.edu/spt/index.php}}, report a small running of the spectral index; $-0.046 < \alpha_s < -0.003$ at $95\%$ confidence \citep{SPT_cos}. However, in general, any such detections have been weak and were consistent with zero.  

\citet{PlanckCP} extensively investigated features in the primordial power spectrum. A penalised likelihood approach indicated that there might be a feature near the highest wavenumbers probed by Planck at an estimated significance of $\sim 3 \sigma$. This nominally statistically significant feature is detected around $k\sim0.13 \mathrm{\;Mpc}^{-1}$. It has been confirmed that the large dip at $\ell\sim1800$ in the CMB power spectrum, which is associated with residual electromagnetic interference generated by the drive electronics of the 4 K cooler, is in fact responsible for the features detected at these high wavenumbers. 

With the recent release of BICEP2 B-mode polarisation data there has been a variety of studies investigating the shape of both the scalar and the tensor primordial power spectrum in light of these new data. \citet{HazShaf_RuleOutPL} have combined Planck CMB temperature and BICEP2 B-mode polarisation data \citep{BICEP2_1,BICEP2_2} to show that, assuming inflationary consistency relation, the power-law form of the scalar primordial spectrum is ruled out at more than $3\sigma$ confidence level. In fact, a break or step at large scales in the primordial scalar perturbation spectrum is more favourable. 

Determining the shape of the primordial spectrum generally consists of two approaches, one by parametrisation, the second by reconstruction. Non-parametric methods suffer from the non-invertibility of the transfer function that describes the {\it transfer} from $P(k)$ to the CMB power spectrum. The dependence on the transfer function has the form
\begin{equation} 
C_{\ell}^{\textrm{th}}=4\pi\int_{0}^{\infty}d\ln k\Delta_{\ell}^{2}(k)P(k)\;,
\label{eq:CMB}
\end{equation} 
where $\ell$ is the angular wavenumber that corresponds to an angular scale via $\ell\sim180^{o}/\theta$ and $\Delta_{\ell}(k)$ is the radiation transfer function holding the cosmological parameters responsible for the evolution of the Universe. Because of the singularity of the transfer function and the limitations on the data from effects such as projection, cosmic variance, instrumental noise, and point sources, a robust algorithm is necessary to an accurately reconstruct the primordial power spectrum from CMB data.

In this letter, we use our recently published algorithm PRISM \citep{PRISM_WMAP9} to reconstruct this spectrum from the LGMCA Planck PR1 data \citep{PR1_WPR1}.

\section{PRISM}
\label{sec:prism}

A CMB experiment, such as Planck, measures the anisotropies in the CMB temperature $\Theta(\vec{p})$ in direction $\vec{p}$, which is described as $T(\vec{p}) = T_{\mathrm{CMB}} [1 + \Theta(\vec{p}) ]$. 
This anisotropy field can be expanded in terms of spherical harmonic functions $Y_{\ell m}$ as
$\Theta(\vec{p}) = \sum_\ell \sum_m a_{\ell m} Y_{\ell m}(\vec{p}),$
where $a_{\ell m}$ are the spherical harmonic coefficients that have a  Gaussian distribution with zero mean, $\langle a_{\ell m} \rangle = 0$, and variance $ \langle a_{\ell m} a^*_{\ell^{\prime} m^{\prime}} \rangle = \delta_{\ell \ell^{\prime}} \delta_{mm^{\prime}} C_\ell^{\textrm{th}}$. 

In practice, we only observe a realisation of this underlying power spectrum on our sky, meaning we are restricted by cosmic variance, especially on large scales. In addition, data are contaminated with additive instrumental noise on small scales and, because of different Galactic foregrounds, some areas of the observed CMB map need to be masked, which induces correlations between modes. Taking these effects into account and following the MASTER method from \citet{MASTER}, the pseudo power spectrum $\widetilde{C}_\ell$ and the empirical power spectrum $\widehat{C}^{\mathrm{th}}_\ell$, which is defined as $\widehat{C}^{\mathrm{th}}_\ell = 1/(2\ell + 1) \sum_m | a_{\ell m} |^2$, can be related through their ensemble averages
\begin{equation}
\langle \widetilde{C}_\ell \rangle = \sum_{\ell^\prime} M_{\ell \ell^\prime}   \langle \widehat{C}^{\mathrm{th}}_{\ell^\prime} \rangle +  \langle \widetilde{N}_\ell \rangle \;,
\end{equation}
where $M_{\ell \ell^\prime}$ describes the mode-mode coupling between modes $\ell$ and $\ell^\prime$ resulting from computing the transform on the masked sky. We note that in this expression $ \langle \widehat{C}^{\mathrm{th}}_{\ell^\prime} \rangle =  C^{\mathrm{th}}_{\ell^\prime} $, and we set $C_\ell = \langle \widetilde{C}_\ell \rangle$ and $N_\ell = \langle \widetilde{N}_\ell \rangle$,
where $C_\ell$ and $N_\ell$ refer to the CMB and the noise power spectra of the masked maps, respectively.
We have assumed that the pseudo power spectrum $\widetilde{C}_\ell$ follows a $\chi^2$ distribution with $2 \ell + 1$ degrees of freedom and hence can be modelled as
\begin{equation}
\widetilde{C}_\ell =\left( \sum_{\ell^\prime} M_{\ell \ell^\prime} C^{\mathrm{th}}_{\ell^\prime} + N_\ell \right) Z_\ell \label{eq:pseudoToTheoPS} \;,
\end{equation}
where $Z_\ell$ is a random variable distributed according to $(2 \ell +1 )Z_{\ell} \sim \chi^2_{2\ell + 1}$.

\subsection{Formulation of the inverse problem}

The relation between the discretised primordial power spectrum $P_{k}$ and the measured pseudo power spectrum $\widetilde{C}_\ell$, computed on a masked noisy map of the sky, can be condensed into the following form:
\begin{equation}
\widetilde{C}_\ell = \left( \sum_{\ell^\prime k } M_{\ell \ell^\prime} T_{\ell^\prime k}P_{k} +   N_\ell \right) Z_\ell\;,
\label{eq:complete-problem-mult}
\end{equation}
with matrix elements $T_{\ell k}=4 \pi \Delta\ln k\,\Delta_{\ell k}^2$, where $\Delta\ln k$ is the logarithmic $k$ interval for the discrete sampling chosen in the integration of the system of equations. Because of the non-invertibility of the $\mathbf{T}$ operator, recovering the primordial power spectrum $P_k$ from the true CMB power spectrum $C_\ell^{\mathrm{th}}$ already constitutes an ill-posed inverse problem, made even more difficult by the mask and cosmic variance affecting the observed $\widetilde{C}_\ell$. In PRISM, we address both the inversion problem and the control of the noise on the CMB spectrum due to sample variance in the framework of sparse recovery. The inversion problem in Equation~\ref{eq:complete-problem-mult} can be regularised in a robust way by using the sparse nature of the reconstructed signal as a prior.

\subsection{$P_{k}$ sparse recovery formulation}

If the signal to recover, $P_k$ in our case, can be sparsely represented in an adapted dictionary $\mathbf{\Phi}$, then this problem, known as the basis pursuit denoising BPDN, can be recast as an optimisation problem. In our case, the optimisation problem can be formulated as
\begin{equation}
\min\limits_X \frac{1}{2} \parallel C_\ell - (\mathbf{M}\mathbf{T} X + N_\ell) \parallel_2^2 + \lambda \parallel \mathbf{\Phi}^t X \parallel_0\;,
\label{eq:bp-lagragian}
\end{equation}
where $X$ is the reconstructed estimate for the primordial power spectrum $P_k$. The first term in Eq. (\ref{eq:bp-lagragian}) imposes an $\ell_2$ fidelity constraint to the data, while the second term promotes the sparsity of the solution in dictionary $\mathbf{\Phi}$. The parameter $\lambda$ tunes the sparsity constraint.

Although the $\ell_0$ optimisation problem stated in Equation~\eqref{eq:bp-lagragian} cannot be solved directly, its solution can be estimated by solving a sequence of weighted $\ell_1$ minimisation problems \citep{Candes2007} of the form
\begin{equation}
\min\limits_X \frac{1}{2} \parallel \frac{1}{\sigma_\ell}\overline{R}_\ell(X) \parallel_2^2 + K \sum_i \lambda_i | [ \mathbf{W} \mathbf{\Phi}^t  X ]_i  |\;,
\label{eq:reweighted-bp-lagragian}
\end{equation}
where $\mathbf{W}$ is a diagonal matrix applying a different weight for each wavelet coefficient, $\overline{R}_\ell(X)$ is an estimate of the residual $C_\ell -(\mathbf{M}\mathbf{T} X + N_\ell)$, and $\sigma_\ell$ is its standard deviation. The parameters $\lambda_i$ are set with respect to the expected standard variation $\sigma(w_i)$ of each wavelet coefficient such that $\lambda_i = K \sigma(w_i)$, where $K$ is a global regularisation parameter (usually set to $K=3$ or $K=4$), which translates into a significance level threshold for the detection of features. The weighted $\ell_1$ problem \eqref{eq:reweighted-bp-lagragian} is solved several times using the popular iterative soft-thresholding algorithm (ISTA) and updating each time the weights $\mathbf{W}$ based on the solution of the previous iteration. This procedure is fully described in \citet{PRISM_WMAP9}.

\section{Results}
\label{sec:results}

Before applying the PRISM algorithm to the Planck data processed with the LGMCA\footnote{LGMCA codes and Planck PR1 data are available at \url{http://www.cosmostat.org/planck_pr1.html}} pipeline, we investigate the performance of the algorithm on a set of simulations. The Planck team provides 100 of their 1000 CMB and instrumental noise simulated maps\footnote{\url{http://wiki.cosmos.esa.int/planckpla}}. The cosmology used to set up these CMB simulations is different from the PR1 best-fit cosmology. Because we wish to have the exact same pipeline applied to both the data and the simulations (hence the same radiation transfer function), we created our own 100 CMB simulations based on PR1 fiducial cosmology. However, we used the simulated noise maps provided by the Planck team. To enable a thorough comparison between the simulations and the Planck PR1 data, we ran the simulations through the same pipeline as the Planck data, that is, through the LGMCA pipeline. For each simulation, we processed the nine frequency channels through LGMCA, with the precomputed set of parameters \citep{PR1_WPR1}. Full-sky noisy maps with a 5 arcmin resolution were obtained, which were then masked using a Galactic and point sources mask with $f_{\textrm{sky}} = 0.76$. The pseudo power spectra were obtained by applying the empirical power spectrum estimator to the masked maps. We also built an estimate of the instrumental noise power spectrum $N_\ell$ by processing the noise maps through the same pipeline in LGMCA, masking the resulting noise maps and applying the pseudo spectrum estimator to these masked maps. We set our estimate of $N_\ell$ to the average of the 100 noise pseudo spectra.

To apply PRISM to the simulated data, we built a transfer function $\mathbf{T}^\prime$ adapted to the simulations, so that it includes the effects of the beam $b_\ell$ and the HEALPix window $h_\ell$ represented by the two diagonal matrices $\mathbf{B} = \mathrm{diag}(b^2_\ell)$ and $\mathbf{H} = \mathrm{diag}(h^2_\ell)$; $\mathbf{T}^\prime = \mathbf{H} \mathbf{B} \mathbf{T} $. The radiation transfer function $\mathbf{T}$ was computed for the PR1 fiducial cosmology using CAMB (\url{http://camb.info}). The lensing contribution to the CMB temperature power spectrum, also computed with CAMB for the PR1 fiducial cosmology, is taken into account as an additional contribution to $N_\ell$.

In contrast to \citet{PRISM_WMAP9}, where the algorithm was initialised to a scale-invariant power spectrum, in this work we initialised the algorithm to the best-fit Planck PR1 primordial power spectrum because we search for small deviations from the best-fit power law that already fits the data. With this choice of initialisation, the reconstruction will not depart from the best-fit power law in the absence of evidence from the data. To reconstruct the simulations and the Planck data, we fixed the regularisation parameter $K$ to $K=4$ which robustly removes the noise due to sampling variance, and we used bi-orthogonal Battle-Lemari\'e wavelets of order 1 with nine dyadic scales. This choice of wavelet dictionary is generic and not specifically tuned to recover physically motivated features, but these wavelets exhibit limited oscillations and have two vanishing moments, which makes them well suited to recover a near scale-invariant power spectrum in logarithmic scale. 

We note that reconstructing the primordial power spectrum is limited by different effects on different scales. On very large scales, the recovery is limited by cosmic variance and geometrical projection of the modes, and on small scales we are limited by effects such as instrumental noise and point sources. These limitations leave us a window through which the primordial power spectrum can be constrained with good accuracy and through which features can be detected. For the Planck data, we expect this window to be in the range $k\sim0.005-0.20 \mathrm{\;Mpc}^{-1}$.

\begin{figure}[t]
\includegraphics[width=\columnwidth]{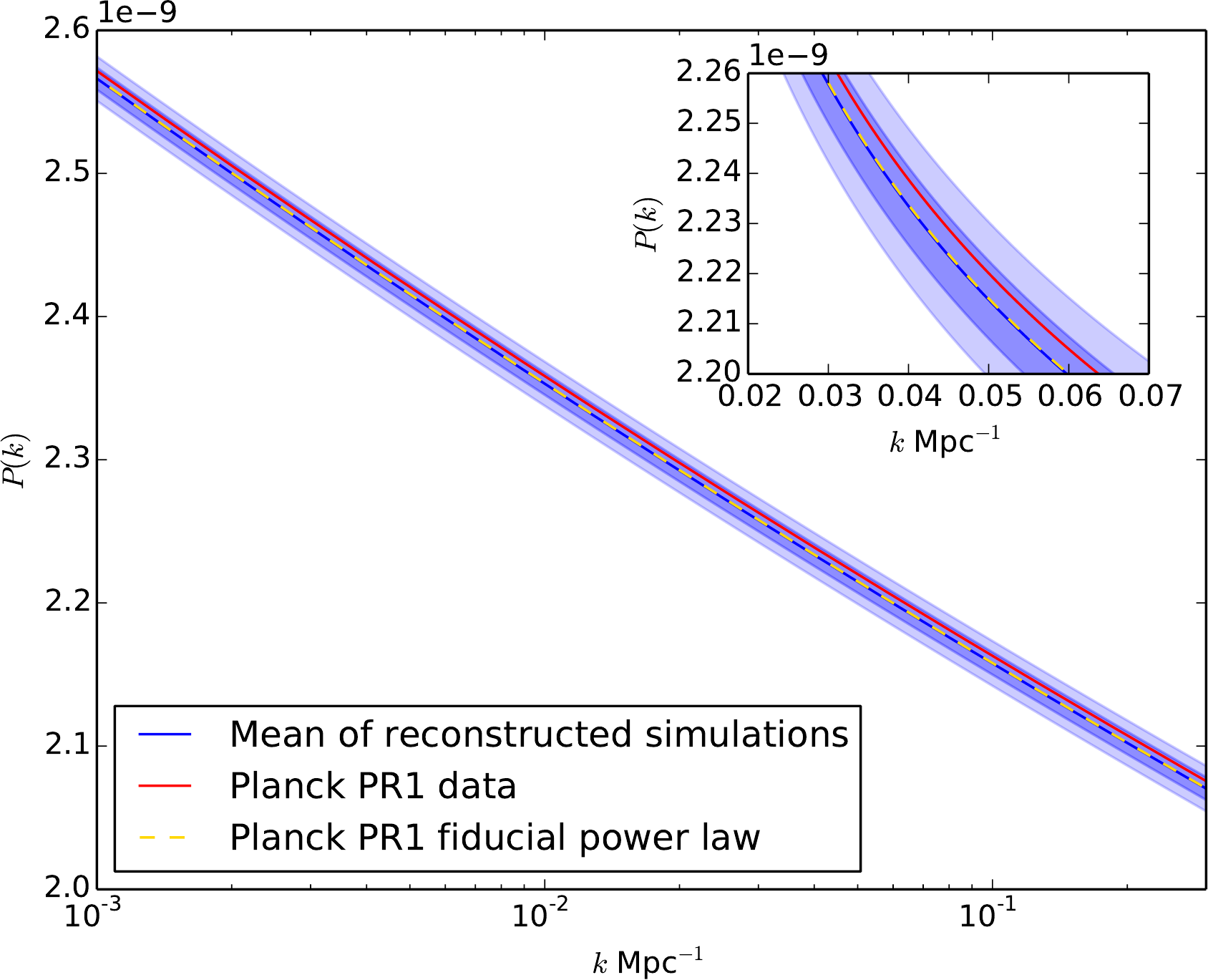}
\caption{\small{Reconstructions for the primordial power spectra from 100 simulations and Planck PR1 data. The $1\sigma$ and $2\sigma$ dispersion of the reconstructed spectra from the simulations are shown as blue bands around the mean of the reconstructions (blue line). We note that these bands do not include the errors due to point sources and beam uncertainties. The Planck fiducial power spectrum used for the simulations, with $A_s = 2.215 \times 10^{-9}$ and $n_s = 0.9626$, is shown as a yellow dashed line. The inset shows a close-up of the main figure.}}
\label{fig:Fig1}
\end{figure}

In Fig.~\ref{fig:Fig1} we show the reconstructed spectra from the simulations and the data. The mean reconstructed power spectrum perfectly fits the input PR1 best-fit power law in the entire reconstructed range. Of course this does not mean that the algorithm is able to perfectly reconstruct an unknown power spectrum over this entire range, but that with the regularisation level used for these reconstructions, no significant departures from the best-fit power law have been detected. The reconstructed spectrum from the LGMCA PR1 power spectrum remains within the $1\sigma$ bar of the reconstructed spectra from the PR1 best-fit power law. Thus, we find no significant departure from the PR1 best-fit near scale-invariant spectrum.

\begin{figure}[t]
\includegraphics[width=\columnwidth]{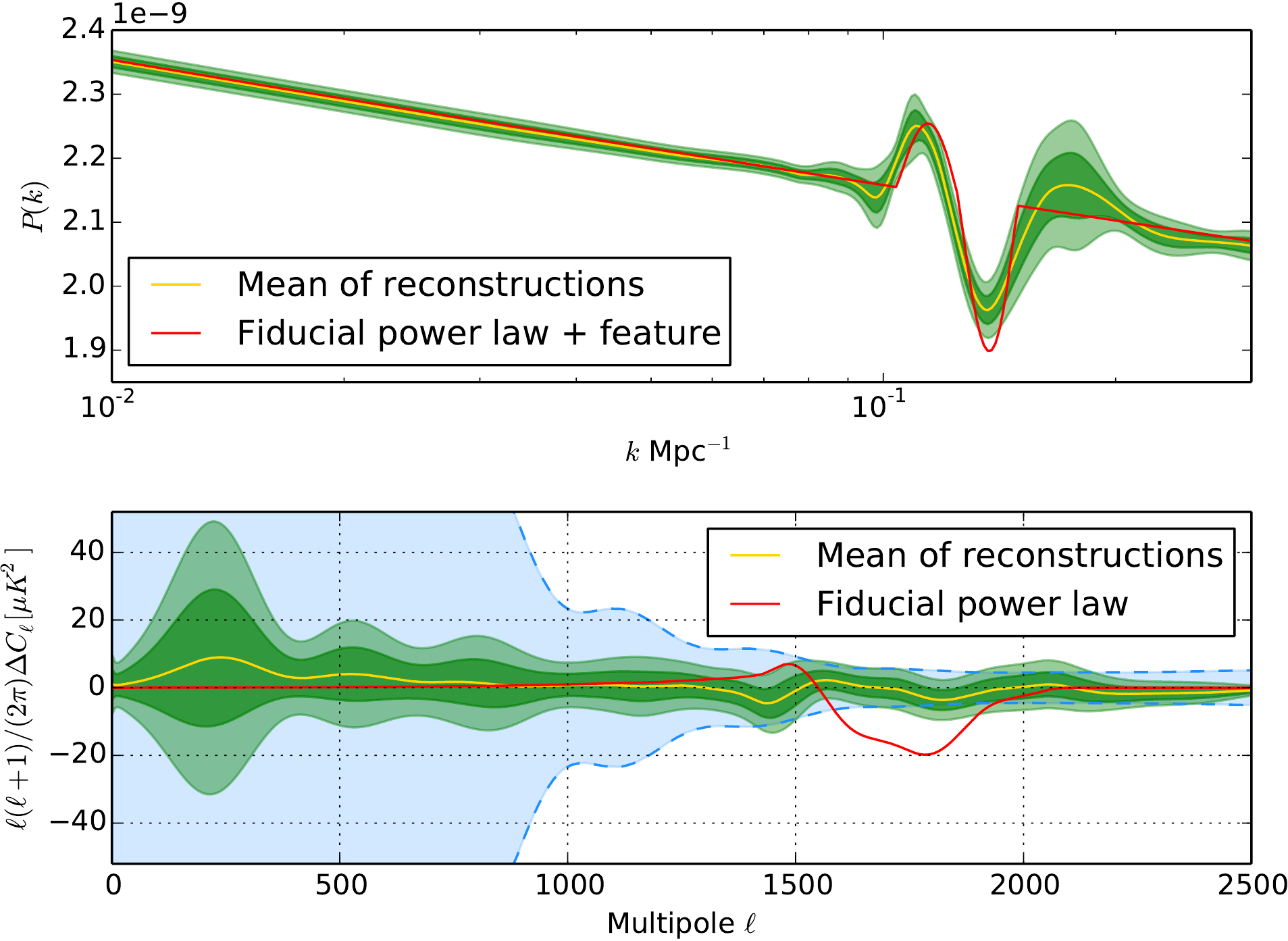}
\caption{\small{Top panel shows a fiducial primordial power spectrum with a feature around $k=0.125$ Mpc$^{-1}$ in red and in green contours the $1\sigma$ and $2\sigma$ dispersion of 100 reconstructions from simulated CMB spectra. The mean of reconstructions is shown in yellow. The bottom panel shows the residuals $\Delta C_\ell$ between the $C_\ell$ for the fiducial primordial power spectrum with a feature and the  $C_\ell$ for the best-fit Planck power law in red and for the mean reconstructed primordial power spectrum in yellow. The green bands indicate the $1\sigma$ and $2\sigma$ bands for the $\Delta C_\ell$ from the simulations, the dashed blue lines show the $1\sigma$ region due to cosmic variance.}}
\label{fig:Fig2}
\end{figure}

As a complementary test of PRISM on Planck-like data we assessed the algorithm's ability to recover a small local departure from the best-fit PR1 power law. We created a set of CMB simulations from a fiducial primordial power spectrum with a small localised test feature causing a dip in the angular power spectrum around $\ell \sim 1800$. The aim of this set of simulations was to mimic the feature that \citet{PlanckCP} proposed to be accountable for the large dip in the angular power spectrum, which was later confirmed as being caused by residual electromagnetic interferences. Our test primordial power spectrum was built from the best-fit PR1 power law with an added feature around $k=0.125$ which causes a dip in the angular power spectrum around $\ell \sim 1800$. This feature and the residuals $\Delta C_\ell$ between the fiducial angular power spectrum and the PR1 best fit $C_\ell$ are shown in Fig.~\ref{fig:Fig2}.

From this test primordial power spectrum, we generated a set of 100 CMB simulations using the exact same procedure as previously mentioned, and we applied PRISM to the measured angular pseudo-power spectra with the exact same parameters. As can be seen in Fig.~\ref{fig:Fig2}, the feature is successfully detected, and the reconstruction shows little bias in position and amplitude. Using the primordial power spectrum reconstructed with PRISM enables a much better fit to the data than a power law, and the reconstructed angular power spectra fall inside the $1\sigma$ region due to cosmic variance. If such a feature existed in the LGMCA processed Planck PR1 data, PRISM would therefore have been able to detect it.

\section{Conclusions}
\label{sec:conclusion}

The simplest inflation models are currently favoured the most  by the data and predict a near scale-invariant power spectrum with a possible small running. The CMB spectrum provides a possibility to measure the primordial spectrum. However, the singular nature of the radiation transfer function, the joint estimation of the cosmological parameters and the primordial power spectrum, along with the different sources of noise, impede the full recovery of the primordial spectrum. Therefore, devising a robust technique able to detect deviations from scale invariance is important.

We have applied our recently published algorithm PRISM \citep{PRISM_WMAP9} to the Planck PR1 data to recover the primordial power spectrum. PRISM is a sparse recovery method, that uses the sparsity of the primordial power spectrum as well as an adapted modelling for the noise of the CMB power spectrum. This algorithm assumes no prior shape for the primordial spectrum and does not require a coarse binning of the power spectrum, making it sensitive to both general smooth features (e.g. running of the spectral index) and local sharp features (e.g. a bump or an oscillatory feature). We reconstructed the primordial power spectrum from the LGMCA PR1 Planck dataset in the range $k\sim0.005-0.20 \mathrm{\;Mpc}^{-1}$. We did not detect any significant deviations from the best-fit Planck near scale-invariant power spectrum with $A_s=2.215\times10^{-9}$ and $n_s = 0.9624$. However, we tested that PRISM would have been able to recover a small localised feature around $k \sim 0.125$, similar to the feature that was proposed in \citet{PlanckCP} to be accountable for a dip in the angular power spectrum at around $\ell \sim 1800$, if it were present on the LGMCA Planck PR1 data.

\begin{acknowledgements}
This work is supported by the European Research Council grant SparseAstro (ERC-228261), and the Swiss National Science Foundation (SNSF).
\end{acknowledgements}

\bibliographystyle{aa} 
\bibliography{biblio}

\end{document}